\documentclass[prd,twocolumn, nofootinbib]{revtex4}

\pdfoutput=1

\usepackage{graphicx}
\usepackage{amssymb,amsmath,bm}  

\usepackage{lipsum}

\usepackage[usenames,dvipsnames]{xcolor}
\usepackage{hyperref}
\hypersetup{
    colorlinks = true,
    citecolor = {MidnightBlue},
    linkcolor = {BrickRed},
    urlcolor = {BrickRed}
}
\usepackage{lipsum}  
\usepackage{hyperref}
\usepackage{fontawesome}

\newcommand{\ba}{\begin{eqnarray}}
\newcommand{\ea}{\end{eqnarray}}

\begin{document}
\title{The T-E correlation coefficient of Planck legacy data}
\author{ Thibaut Louis$^1$, Zack Li$^2$, Matthieu Tristram$^1$
             }
             \affiliation{$^{1}$LAL, Univ. Paris-Sud, CNRS/IN2P3, Universit\'e Paris-Saclay, Orsay, France\\
             		   $^{2}$  Department of Astrophysical Sciences, Princeton University, Princeton, NJ 08544, USA}
             
\begin{abstract}
Testing deviations from the $\Lambda$CDM model using the Cosmic Microwave Background (CMB) power spectra requires a pristine understanding of instrumental systematics. In this work we discuss the properties of a new observable ${\cal R}^{TE}_{\ell}$, the correlation coefficient of temperature and E modes. We find that this observable is mostly unaffected by systematics introducing multiplicative biases such as errors in calibration, polarisation efficiency, beam and transfer function measurements. We discuss the dependency of this observable on the cosmological model and derive its statistical properties. We then compute the T-E correlation coefficients of Planck legacy data and compare them with expectations from the Planck best-fit $\Lambda$CDM and foreground model.
\end{abstract}

  \date{\today}
  \maketitle

\section{Introduction}\label{sec:intro}

Recent results from the cosmic distance ladder \cite{2019ApJ...876...85R} and  time delays of gravitationally lensed quasars  \cite{2019arXiv190704869W} challenge the $\Lambda$CDM concordance model.  The $H_{0}$ value inferred from the combination of these two low-redshift probes is $5 \sigma$ discrepant with the one inferred from the measurement of the cosmic microwave background by the Planck satellite \cite{2018arXiv180706205P,2018arXiv180706209P}. 
This discrepancy could reveal new physics beyond the standard model or be due to unmodeled systematics. 

A discussion is on-going on the possible systematic effects that could affect the low-redshift observations, in particular the calibration of the cosmic distance ladder (e.g \cite{2019arXiv190705922F}, \cite{2018arXiv180603849R}). 
Models of new physics beyond $\Lambda$CDM  that would reconcile the two measurements have also been developed (see \cite{2019arXiv190803663K} for a review). A particularly interesting class of models proposes to change the acoustic scale by changing physics near the time of recombination \cite{2019PhRvL.122v1301P, 2019arXiv190402625P}.
A general feature of such solutions is a residual to fits to  $\Lambda$CDM that could in principle be detectable by the next generation of CMB experiments. The magnitude of these residuals is however extremely small and measuring them will require very accurate observations. 

The most constraining summary statistics of CMB fluctuations are the power spectra of temperature and E modes anisotropies: $C^{TT}_{\ell},C^{TE}_{\ell},C^{EE}_{\ell}$. The measurement of these spectra requires a complete understanding of the instrument model. The beam of the telescope, its absolute calibration and polarisation efficiency need to be measured.  Transfer functions, due to the filtering of the time-ordered-data or imperfection of the map maker, can also affect the spectra and their effects need to be characterised.

One way to assess the impact of systematics arising from an incomplete instrument model is to compare results from different CMB experiments  \cite{2017JCAP...06..031L,2014JCAP...07..016L,2017ApJ...850..101A,2015ApJ...801....9L,2018ApJ...869...38H}. The forthcoming Simons Observatory \cite{2019JCAP...02..056A} experiment  covering half the sky at high angular resolution will  provide pristine verification of Planck cosmology \cite{2019PhRvD.100b3518L}.

Another complementary approach is to look for CMB observables that are less sensitive on the instrument model.
In this work, we discuss the properties of one of these observables, the correlation coefficient ${\cal R}^{TE}_{\ell}= C^{TE}_{\ell}/\sqrt{C^{TT}_{\ell}C^{EE}_{\ell}}$. This ratio is insensitive to any systematics that takes the form of a multiplicative bias in temperature or E modes. It does however contains information on cosmological parameters and can therefore be used as a robust consistency test for the  $\Lambda$CDM model. 

A precise estimate of ${\cal R}^{TE}_{\ell}$ requires a precise measurement of  E modes. It is therefore an observable suitable for current CMB experiments such as ACTPol \cite{2017JCAP...06..031L}, SPTpol \cite{2018ApJ...852...97H}, Simons Array \cite{2016JLTP..184..805S} and Planck \cite{2018arXiv180706205P,2018arXiv180706209P}, and for the next generation CMB telescopes such as Simons Observatory \cite{2019JCAP...02..056A} and CMB S4 \cite{2016arXiv161002743A}.

This paper is structured as follows. In Section \ref{sec:definition} we introduce the correlation coefficient and display its dependency on the cosmological model. In Section \ref{sec:stat_properties} we discuss the statistical properties of an estimator for ${\cal R}^{TE}_{\ell}$. In Section \ref{sec:Planck}, we compute the correlation coefficient of the Planck legacy data and compare it to $\Lambda$CDM predictions. We conclude in Section \ref{sec:discussion}.

We make the entire code used for this analysis public, it includes tools for computing the correlation coefficient of Planck data,  generating simplified simulations of these data and analysing them. It also contains scripts for reproducing all the figures of this paper. It is available and documented at  \href{https://github.com/simonsobs/PSpipe/tree/master/project/correlation_coeff}{\faGithub}.

\section{Correlation coefficient}\label{sec:definition}

In this section, we introduce the correlation coefficient ${\cal R}^{TE}_{\ell}$. We show that it is robust against systematics leading to multiplicative biases. We then discuss how it varies  while changing  cosmological parameters.

\begin{figure*}
  \centering
  \includegraphics[width=1\textwidth]{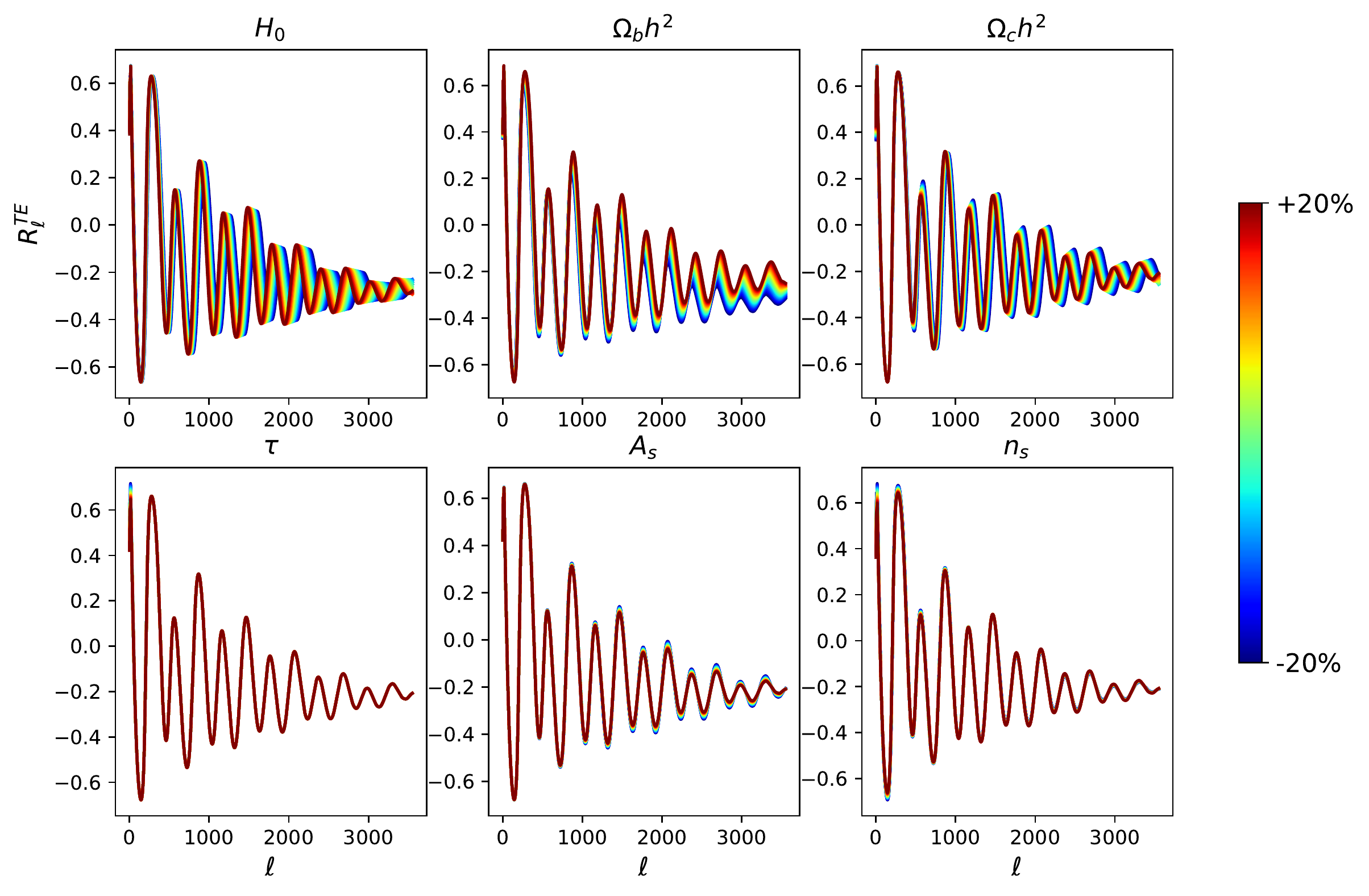}
  \caption{Variations of the correlation coefficient as a function of the cosmological model. We choose central values of parameters compatible with Planck legacy cosmology \cite{2018arXiv180706205P,2018arXiv180706209P} and display a $ \pm 20\%$ variation. The correlation coefficient is poorly sensitive to parameters affecting the overall shape or amplitude of the spectra, but is very sensitive to parameters changing the peaks positions. By definition the correlation coefficient takes values between -1 and 1.  }
  \label{fig:cosmo_dependency}
\end{figure*}

\subsection{Robustness}\label{sec:robust}

A simplified model of the measured temperature and E modes power spectra with respect to the true power spectra can be written
\ba
C^{\rm TT,obs}_{\ell} &=& c_{T}^{2}(b^{T}_{\ell})^{2} (F^{T}_{\ell})^{2} C^{TT}_{\ell}, \nonumber \\
C^{\rm TE,obs}_{\ell} &=& c^{2}_{T} \epsilon b^{T}_{\ell} b^{E}_{\ell}  F^{T}_{\ell}F^{E}_{\ell} C^{TE}_{\ell}, \nonumber \\
C^{\rm EE,obs}_{\ell} &=& c_{T}^{2} \epsilon^{2} (b^{E}_{\ell})^{2} (F^{E}_{\ell})^{2} C^{EE}_{\ell}.
\ea

$b^{T}_{\ell}$ and $b^{E}_{\ell}$ are the azimuthally symmetric harmonic transform of the telescope beams, they account for the finite angular resolution of the telescope. We allow for different beams in temperature and E modes as the optical response of the telescope can vary for intensity and polarisation measurements.
$ c_{T}$ and $\epsilon $ are the absolute calibration and polarisation efficiency respectively, they rescale the overall amplitude of the fluctuations.
 $F^{T}_{\ell}$ and $F^{E}_{\ell}$ are generic transfer functions, while transfer functions can be unity for maximum likelihood map making, they typically arise from un-modeled cut or filtering of the time order ordered data in the map making process. The correlation coefficient ${\cal R}^{TE}_{\ell}= C^{TE}_{\ell}/\sqrt{C^{TT}_{\ell}C^{EE}_{\ell}}$ is unaffected by any of these effects: ${\cal R}^{\rm TE,obs}_{\ell}= {\cal R}^{TE}_{\ell}$.

We note however that ${\cal R}^{TE}_{\ell}$ will be impacted by systematics leading to additive biases, one example is the uncorrected temperature to polarisation leakage in Planck polarisation maps \cite{2018arXiv180706209P,2017A&A...598A..25H}.

\subsection{Dependency on cosmological parameters}\label{sec:cosmo_param}

We display the variation of the correlation coefficient as a function of the six $\Lambda$CDM parameters in Figure \ref{fig:cosmo_dependency}. 
We adopt a  cosmology with    $ \omega_{b}=\Omega_{b}h^{2}= 0.02237$, $ \omega_{c}=\Omega_{c}h^{2}= 0.1200$, $n_{s} =0.9649$, $H_{0}=67.36  \ {\rm km.s}^{-1}\rm{.Mpc}^{-1}$, $\ln(10^{10}A_s)= 3.044$ and $\tau=  0.0544$ compatible with \cite{2018arXiv180706209P}, and vary parameters within $\pm 20\%$ of their fiducial values using the CAMB Boltzmann solver \cite{Lewis:1999bs}. We choose to show the variation of the lensed correlation coefficient, that is, the correlation coefficient formed from the lensed power spectra.

As expected the correlation coefficient is poorly sensitive to parameters affecting the overall shape or amplitude of the spectra, but is very sensitive to parameters changing the peaks positions, in particular the Hubble constant $H_{0}$.  $ \omega_{c}$ also changes the relative amplitude of the peaks and $ \omega_{b}$ changes the overall degree of correlation between temperature and E modes on small scales. One interesting observation is that despite having no information on the overall calibration of the spectra, a parameter such as $A_{s}$ can still be constrained via the smoothing of the acoustic peaks, this is expected as the lensing amplitude will be directly related to the amplitude of fluctuations. Finally, $\tau$ affects the correlation coefficient only on the largest scales related to the reionisation bump.

\section{Statistical properties}\label{sec:stat_properties}
In this section, we discuss in details the estimation of the correlation coefficient and derive analytical expressions for its variance and bias.

\subsection{Estimator}

\begin{figure*}
  \centering
  \includegraphics[width=1\textwidth]{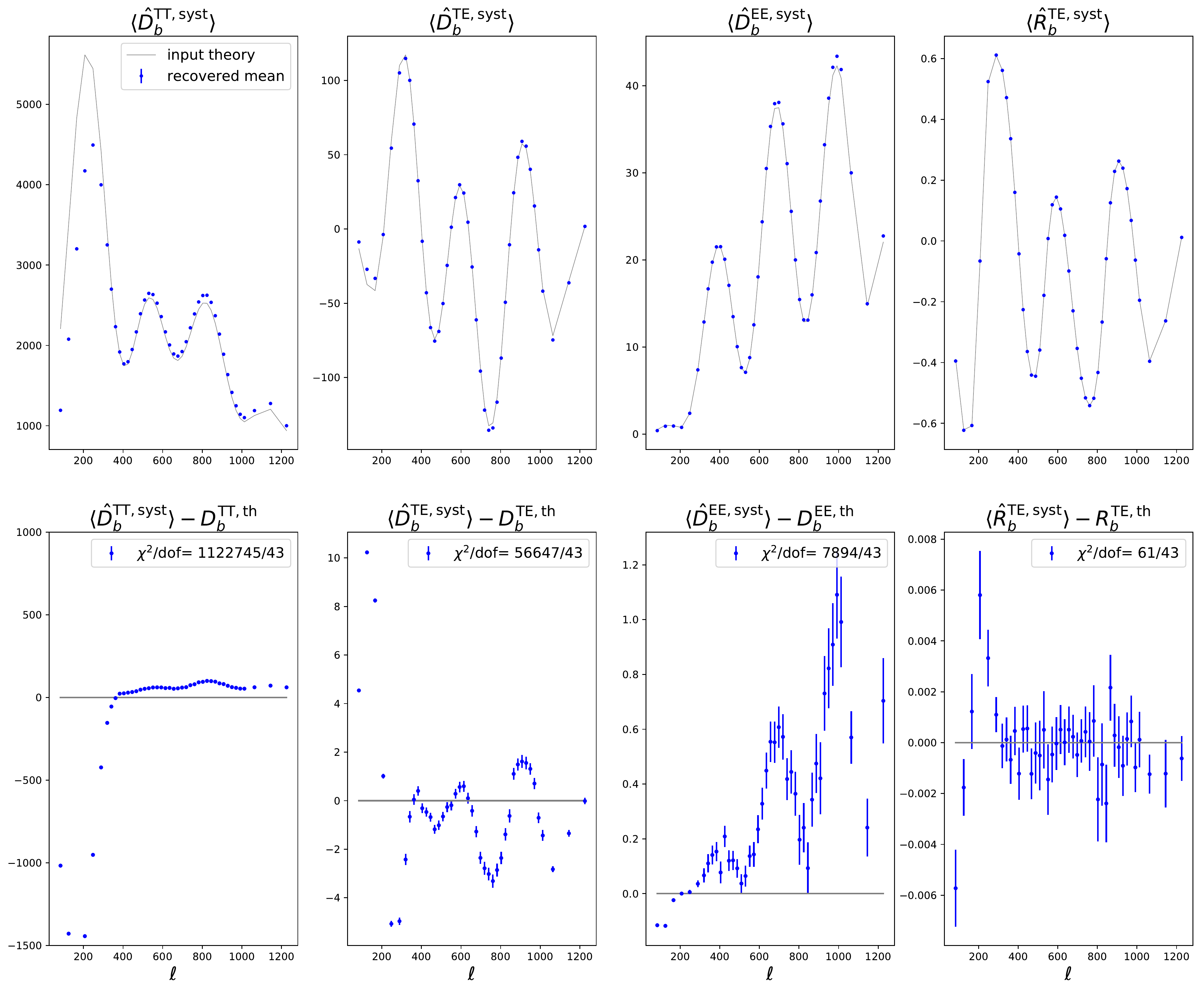}
  \caption{ Impact of beam and transfer function systematics on the recovered mean power spectra of 200 simulations, unlike $\hat{C}^{TT}_{b}$,$\hat{C}^{TE}_{b}$ and $\hat{C}^{EE}_{b}$ the estimated cross correlation coefficient $\hat{R}^{TE}_{b}$ is mostly un-affected by multiplicative biases. The $\chi^{2}$ are computed using monte-carlo errorbar $\sigma_{\rm MC}=\hat{\sigma}/\sqrt{N_{\rm sims}}$. The gray line in the bottom panels represents zero. }
  \label{fig:robustness}
\end{figure*}

We start with the standard estimates of cross power spectra for $N$ splits of data covering the full CMB sky
\ba
\tilde{a}^{i,X}_{\ell m} &=& f^{X}_{\ell}(b^{X}_{\ell} {a}^{X}_{\ell m} + n^{X, i}_{\ell m}), \nonumber \\
\tilde{C}_{\ell}^{XY} &=& \frac{1}{2\ell+1}  \frac{1}{N_{c}} \sum^{N}_{i,j} \sum_{m} \tilde{a}^{i,X}_{\ell m} \tilde{a}^{j,Y *}_{\ell m} (1-\delta_{ij}), \nonumber \\
\hat{C}_{\ell}^{XY} &=& \frac{1}{f^{X}_{\ell} f^{Y}_{\ell} b^{X}_{\ell}b^{Y}_{\ell}  } \tilde{C}_{\ell}^{XY}.
\ea
Here $N_{c}= \sum^{N}_{i,j}(1-\delta_{ij})$ is the number of cross spectra formed from the different splits. X and Y represent either temperature or E modes. $b^{X}_{\ell}$ is the harmonic beam transform and $ f^{X}_{\ell}$ is a general transfer function, also including effects from calibration and polarisation efficiency.
One advantage of using cross power spectra is that there are insensitive to noise biases as long as the splits used to form them have uncorrelated noise $ \langle n^{X, i}_{\ell m}  n^{Y, j *}_{\ell m}\rangle=0$. However in the limit of low signal-to-noise, these cross power spectra can become negative. In order to have a well defined correlation coefficient, we have to bin the power spectra estimates with a bin size chosen so that each bin has a large enough signal-to-noise, $\hat{C}_{b}^{XY}= P_{b\ell} \hat{C}_{\ell}^{XY}$. Another complication comes from the fact that realistic surveys do not have access to the entire CMB sky, the observed map is related to the full sky map by a window function $W(\hat{n})$ leading to a non zero coupling of the observed $a_{\ell m}$s, and a mixing between E and B modes, these effects can be computed and corrected using a mode coupling matrix $M^{\rm XYWZ}_{\ell \ell'}$ \cite{2002ApJ...567....2H,2005MNRAS.360.1262B,2003ApJS..148..161K},  the estimator in its full form is therefore given by
 \begin{widetext}
\ba
\hat{{\cal R}}^{\rm TE}_{b}=\frac{P_{b\ell} (M^{-1})^{\rm TE,TE}_{\ell \ell'} \tilde{C}_{\ell'}^{TE}  }{\sqrt{ \left( P_{b\ell}  (M^{-1})^{\rm TT,TT}_{\ell \ell'} \tilde{C}_{\ell'}^{TT} \right) \left( P_{b\ell} \left[ (M^{-1})^{\rm EE,EE}_{\ell \ell'} \tilde{C}_{\ell'}^{EE}+(M^{-1})^{\rm EE,BB}_{\ell \ell'} \tilde{C}_{\ell'}^{BB} \right]  \right)}} .
\ea
\end{widetext}
We note that with these complications the exact cancellation of systematics due to mis-calibrated beam or transfer function is not guaranteed. In order to test this estimator, we generate 200 simplified simulations of the Planck 143 GHz survey, with white noise levels  $\sigma_{\rm T}=33 \mu \rm{K.arcmin}$, $\sigma_{\rm P}=70.2 \mu \rm{K. arcmin}$, and mask them using the 143 GHz Planck legacy likelihood masks \cite{2018arXiv180706205P,2019arXiv190712875P}. The fiducial beam full width at half maximum (FWHM) of Planck at 143 GHz is $7.30$ arcmin. In order to mimic a systematic beam error, we convolve the simulations with a beam FWHM $5\%$ smaller and deconvolve the fiducial beam. We also include an un-modelled transfer function in the simulations, we parametrise it as:
\ba
f^{X}_{\ell} &=& f^{X}_{\ell=2}+ (1- f^{X}_{\ell=2}) \cos \left( \frac{( \ell_{*}^{X}-\ell)}{ ( \ell_{*}^{X}-2) }\frac{\pi}{2} \right)^{2}
\ea
for $2 \leq \ell< \ell_{*}^{X}$ and 1 otherwise. This function increases smoothly from $ f^{X}_{\ell=2}$ and is equal to one when $\ell=\ell_{*}^{X} $.  We take $f^{T}_{2}=0.7 , f^{E}_{2}=0.8, \ell_{*}^{T}=400,\ell_{*}^{E}=200$. We note that these choices are arbitrary and are only made to illustrate the robustness of the correlation coefficient. In Figure \ref{fig:robustness}, we show the recovered mean power spectra of the simulations and compare it with the input binned theory curves. As expected, systematics have a strong impact on the recovered power spectra of the simulations $\hat{D}^{TT}_{b} $, $\hat{D}^{TE}_{b} $, $\hat{D}^{EE}_{b} $ but do not affect significantly the estimated cross correlation coefficient $\hat{{\cal R}}^{TE}_{b}$, illustrating the robustness of the coefficient against instrumental systematics introducing multiplicative biases such as mis-modeled beams or unknown transfer functions. We also report the $\chi_{\rm MC}^{2}$ of the spectra computed using Monte-Carlo errorbars: $\sigma_{\rm MC}=\hat{\sigma}/\sqrt{N_{\rm sims}}$. The interpretation of  $\chi_{\rm MC}^{2}$ is the following, even if we were to have access to observations $\sim15 \times$ more constraining than Planck 143 GHz, detecting effects from multiplicative systematics on the correlation coefficient would be challenging.

We note that, for illustration purpose, we have exaggerated the impact of systematics. In practice, beams can be precisely estimated from planet measurements, and transfer functions can be assessed using end-to-end simulations, the associated errors on the measurement of these quantities can then be included in the covariance matrix of the spectra. Using the correlation coefficient is however extremely robust against the unknown-unknowns, effects that are not properly modelled in standard analysis of CMB data. 

\begin{figure*}
  \centering
  \includegraphics[width=1\textwidth]{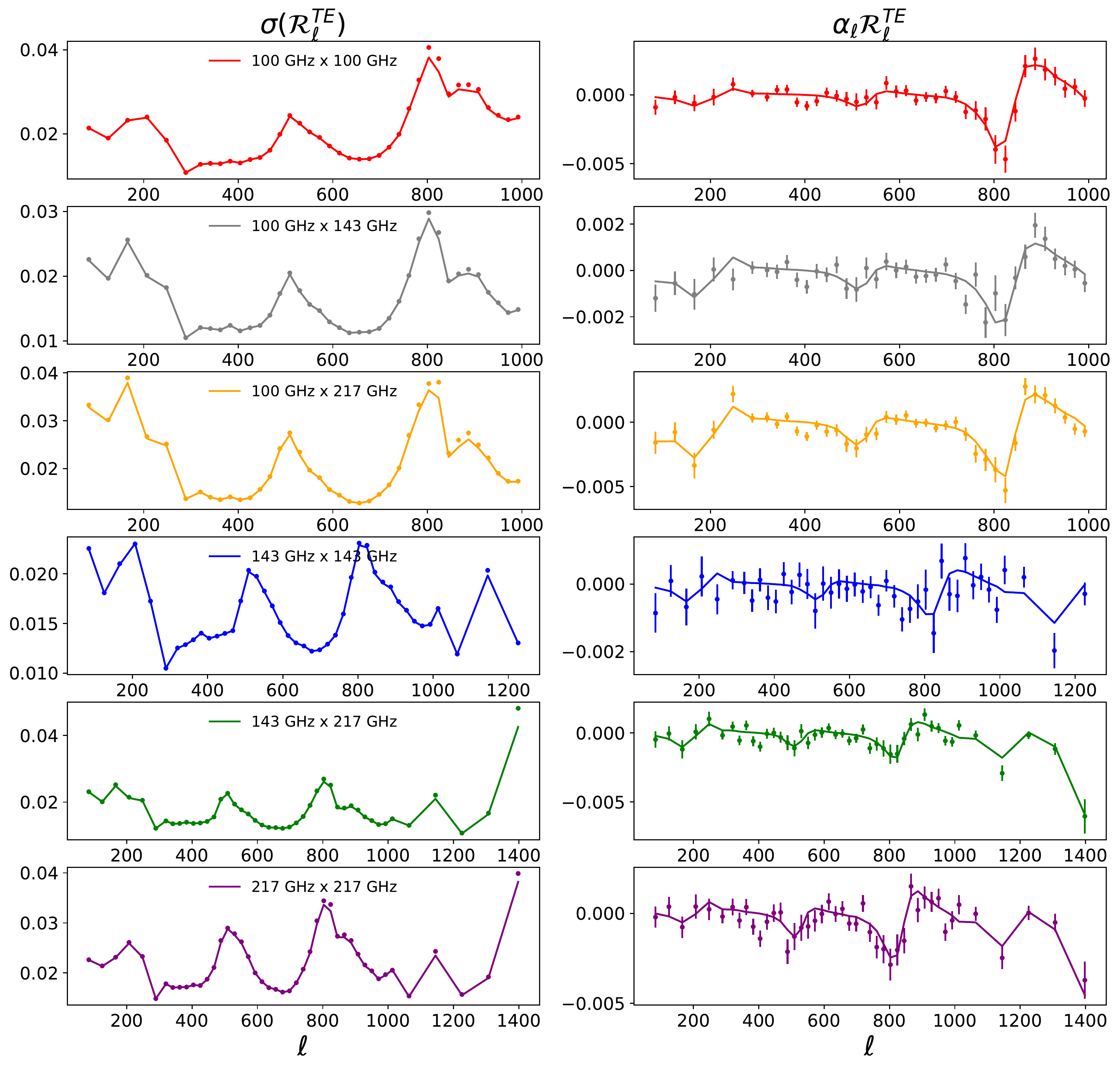}
  \caption{Standard deviations and biases for the 100 GHz, 143 GHz and 217 GHz Planck correlation coefficients also including cross frequency spectra. We compare the bias and dispersion of the correlation coefficient of the simulations (dotted) with our analytic expressions relating them to the $C^{TT}_{\ell},C_{\ell}^{EE},C_{\ell}^{TE}$ covariance matrix elements. The bias is extremely small or order  half a percent of ${\cal R}^{\rm TE}_{b}$ and measuring it has required running 1500 simulations of the Planck survey. The displayed errorbars on the bias are therefore $\sqrt{1500} \approx 38$ times smaller than the real ${\cal R}^{TE}_{\ell}$ errors.  We find a satisfying agreement between the expectation for the statistical properties of ${\cal R}^{TE}_{\ell}$ and the simulations.  }
  \label{fig:cov_and_bias}
\end{figure*}

\subsection{${\cal R}^{TE}_{\ell}$ Covariance}\label{subsec:r_covariance}

To compute the covariance of the estimator of the correlation coefficient, we start by writing a simplified likelihood of CMB data:
\ba
-2 \ln {\cal L} = & \sum_{\ell m \ell' m'} & (a^{T}_{\ell m} \ a^{E}_{\ell m}) C_{\ell m \ell' m'}^{-1} (a^{T}_{\ell' m'} \ a^{E}_{\ell' m'})^{\dagger } \nonumber \\
 &+&\ln ({\rm det} (C))+ {\rm cst}.
\ea
 $C_{\ell m \ell' m'} = C_{\ell} \delta_{ \ell \ell'} \delta_{ m m'}$  is the covariance matrix of the $a_{\ell m}$s, it is given in term of signal and effective noise power spectra
 \ba
C_{\ell}= \begin{pmatrix} 
C^{TT}_{\ell} +N^{\rm TT, eff}_{\ell} &  {\cal R}^{TE}_{\ell}\sqrt{C^{TT}_{\ell}C_{\ell}^{EE}} \\
{\cal R}^{TE}_{\ell}\sqrt{C^{TT}_{\ell}C_{\ell}^{EE}} & C_{\ell}^{EE} + N^{\rm EE, eff}_{\ell}
\end{pmatrix}.
\ea
The effective noise power spectra are defined as the ratio of the noise power spectra to the beams window function $N^{\rm XX, eff}_{\ell}= N^{\rm XX}_{\ell}/(b^{X}_{\ell})^{2}$.
The Fisher matrix associated to this likelihood is given by
\ba
F_{\alpha \beta} = - \left\langle \frac{\partial^{2}  \ln {\cal L}}{\partial \lambda_{\alpha}\partial \lambda_{\beta} } \right\rangle= \frac{1}{2} {\rm Tr} [ C_{,\lambda_{\alpha}} C^{-1} C_{,\lambda_{\beta}} C^{-1} ] .
\ea
The inverse of the fisher matrix is the covariance matrix of the parameters
${\rm Cov}(\lambda_{\alpha})= (F^{-1})_{\alpha \alpha}$.
Using $\lambda= \{ C^{TT}_{\ell},{\cal R}^{TE}_{\ell}, C^{EE}_{\ell}\}$, we obtain an expression for the covariance of the correlation coefficient  ${\rm Cov}({\cal R}^{TE}_{\ell})$
 \begin{widetext}
\ba
\Delta \ell_{b}f_{\rm sky}(2\ell_{b}+1){\rm Cov}({\cal R}^{TE}_{b})  &=& ({\cal R}^{TE}_{b})^{4}-2({\cal R}^{TE}_{b})^{2}+1 + \frac{N^{\rm TT, eff}_{b}}{C^{TT}_{b}}+ \frac{N^{\rm EE, eff}_{b}}{C^{EE}_{b}}  + \frac{ N^{\rm TT, eff}_{b} N^{\rm EE, eff}_{b}}{C^{TT}_{b}C^{EE}_{b}}  \nonumber \\
&+&  ({\cal R}^{TE}_{b})^{2}\left(  \frac{1}{2} \left( \frac{N^{\rm TT, eff}_{b}}{C^{TT}_{b}} -1 \right)^{2}  +  \frac{1}{2} \left( \frac{N^{\rm EE, eff}_{b}}{C^{EE}_{b}} -1 \right)^{2} -1 \right),\label{eq:Cov}
\ea
\end{widetext}
where we introduce the fraction of observed sky $f_{\rm sky}$ to rescale the number of modes available for an experiment observing only part of the sky, and $\Delta \ell_{b}$ accounting for the bin size of the estimator. 
This expression can be re-expressed purely in term of the covariance of the temperature and E modes power spectra
\ba
\frac{{\rm Cov}({\cal R}^{TE}_{b})}{({\cal R}^{TE}_{b})^{2}} &=&   \frac{ {\rm Cov} (C^{TE}_{b })}{(C^{TE}_{b})^{2}} + \frac{1}{4} \left(\frac{ {\rm Cov} (C^{TT}_{b })}{(C^{TT}_{b})^{2}} + \frac{ {\rm Cov} (C^{EE}_{b })}{(C^{EE}_{b})^{2}}  \right)\nonumber \\
 &-& \left(   \frac{ {\rm Cov} (C^{TT}_{b }, C^{TE}_{b } )}{C^{TT}_{b}C^{TE}_{b}}   +  \frac{ {\rm Cov} (C^{EE}_{b }, C^{TE}_{b } )}{C^{EE}_{b}C^{TE}_{b}} \right) \nonumber \\
 &+&\frac{1}{2} \frac{ {\rm Cov} (C^{TT}_{b }, C^{EE}_{b } )}{C^{TT}_{b}C^{EE}_{b}}. \label{eq:equivalence}
\ea
We note that this last expression is especially useful. Since it means that computing the variance of the correlation coefficient does not require any extra computation once the covariance matrix of the $C^{\rm TT}_{\ell},C^{\rm EE}_{\ell},C^{\rm TE}_{\ell}$ spectra is known. An implicit assumption in the derivation of Eq \ref{eq:Cov} is that the effective noise power spectra $N^{\rm XX, eff}_{b}$ are known, it is therefore the correct expression for the variance of  auto-power spectra with a known noise bias subtraction. Eq \ref{eq:equivalence} is however general and can be used both for cross spectra and auto power spectra. 

Others interesting quantities are the joint covariances of ${\cal R}_{\ell}^{TE}$ with the temperature and E modes power spectra. These terms would be important if we were to fit cosmological parameters from the three power spectra. We obtain
\ba
\frac{{\rm Cov}({\cal R}^{TE}_{b}, C^{\rm XX}_{b})}{{\cal R}^{TE}_{b}} &=& \frac{{\rm Cov}(C^{TE}_{b}, C^{\rm XX}_{b})}{C^{TE}_{b}}  -\frac{1}{2} \frac{{\rm Cov}(C^{\rm XX}_{b}, C^{EE}_{b})}{ C^{EE}_{b}} \nonumber \\
&-&\frac{1}{2} \frac{{\rm Cov}(C^{\rm XX}_{b}, C^{TT}_{b}) }{ C^{TT}_{b}},
\ea 
where XX stands for either TT or EE.

\subsection{Bias}

Despite being formed using unbiased cross power spectra, the estimator for the correlation coefficient is biased. To derive an expression of the bias, let us expand the estimator in the limit of high signal to noise: \begin{widetext}
\ba
\hat{{\cal R}}^{TE}_{b}&=& \frac{\hat{C}^{TE}_{b}}{\sqrt{\hat{C}^{TT}_{b}\hat{C}^{EE}_{b}}}= {\cal R}^{TE}_{b}  \frac{ \left(1 + \frac{ \Delta C^{TE}_{b} }{C^{TE}_{b}} \right) }{\sqrt{ \left(1+ \frac{\Delta C^{TT}_{b}}{C^{TT}_{b}} \right) \left(1+\frac{\Delta C^{EE}_{b}}{C^{EE}_{b}} \right) }}  \nonumber \\
&=& {\cal R}^{TE}_{b} \left(1 + \frac{ \Delta C^{TE}_{b} }{C^{TE}_{b}} \right)\left(1- \frac{1}{2} \frac{\Delta C^{TT}_{b}}{C^{TT}_{b}} + \frac{3}{8} \left(\frac{\Delta C^{TT}_{b}}{C^{TT}_{b}} \right)^{2} + ... \right) \left(1- \frac{1}{2} \frac{\Delta C^{EE}_{b}}{C^{EE}_{b}} + \frac{3}{8} \left(\frac{\Delta C^{EE}_{b}}{C^{EE}_{b}} \right)^{2} + ... \right). \nonumber \\ \label{eq:development}
\ea
\end{widetext}
An approximate expression for the bias can be obtained by taking the expectation value of this expression. All linear term vanish and keeping the second order term we obtain
$\langle  \hat{{\cal R}}^{TE}_{b} \rangle= {\cal R}^{TE}_{b} ( 1+ \alpha_{b}) $.
With $\alpha_{b}$ given as a function of the covariance matrix elements
\ba
\alpha_{b} &=& \frac{3}{8} \left(\frac{ {\rm Cov} (C^{TT}_{b })}{(C^{TT}_{b})^{2}}+  \frac{ {\rm Cov} (C^{EE}_{b })}{(C^{EE}_{b})^{2}} \right)\nonumber \\ 
 &-&   \frac{1}{2} \left(  \frac{ {\rm Cov} (C^{TT}_{b }, C^{TE}_{b } )}{C^{TT}_{b}C^{TE}_{b}} +  \frac{ {\rm Cov} (C^{EE}_{b }, C^{TE}_{b } )}{C^{EE}_{b}C^{TE}_{b}} \right) \nonumber \\ 
&+&  \frac{1}{4}\frac{ {\rm Cov} (C^{TT}_{b }, C^{EE}_{b } )}{C^{TT}_{b}C^{EE}_{b}}.
\ea
We note that for noiseless data, this expression simplifies to a standard result for bias in estimate of correlation coefficient \cite{bias}
\ba
\langle  \tilde{{\cal R}}^{TE}_{b} \rangle=  {\cal R}^{TE}_{b} \left(1 + \frac{ ({\cal R}^{TE}_{b})^{2 } -1 }{2 \nu_{b}} \right),
\ea
where $\nu_{b}=(2\ell_{b}+1)f_{\rm sky} \Delta_{\ell_{b}}$ is the number of modes used for estimating ${\cal R}^{TE}_{b}$.
In practice, this bias is  subdominant. A bias corrected estimator should nonetheless be formed as
\ba
\hat{{\cal R}}^{TE, \rm c}_{b}= \hat{{\cal R}}^{TE}_{b}(1-\alpha_{b}).
\ea
For Planck legacy data, $\alpha_{b}$ is of order $0.5\%$, the remaining bias of this corrected estimator, of order ${\cal O}(\alpha^{2}_{b})$,  is therefore negligible for all practical purposes.
\begin{figure*}
  \centering
  \includegraphics[width=1\textwidth]{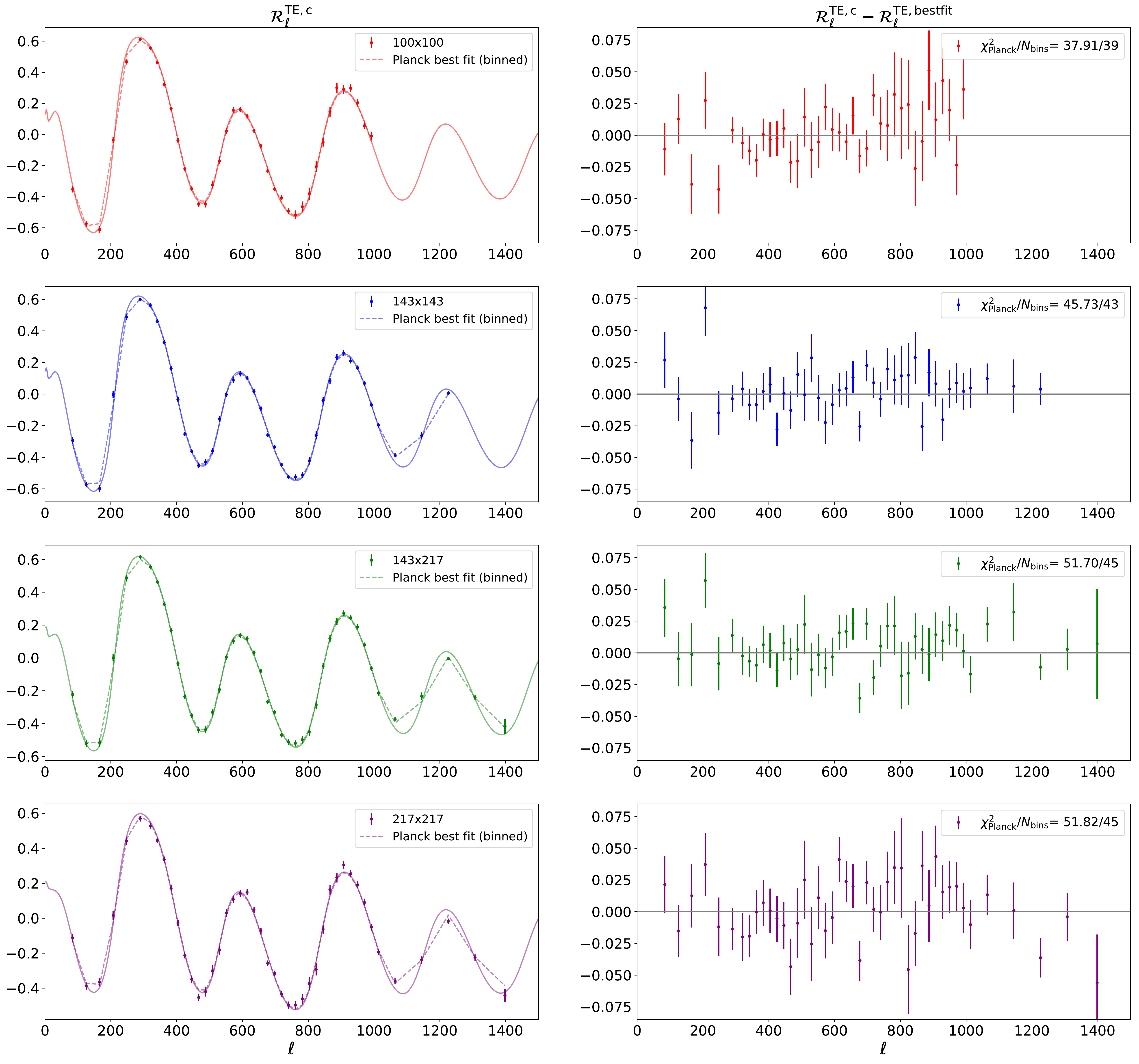}
  \caption{Planck legacy data correlation coefficients for  cross-frequency cross half-mission spectra. The left panels display the correlation coefficient and the best-fit LCDM model, including foreground and systematics. The plik likelihood only models  the 100 GHz $\times$ 100 GHz, 143 GHz $\times$ 143 GHz, 143 GHz $\times$  217 GHz  and 217 GHz $\times$ 217 GHz spectra so we focus on these spectra in this figure.  The right panels show the residuals with respect to the best-fit plik cosmology including systematic and foregrounds. We find no clear features in the residuals. We make the resulting correlation coefficients public, they can be accessed at 
\href{https://github.com/simonsobs/PSpipe/tree/master/project/correlation_coeff/results}{\faGithub  }. }
  \label{fig:results}
\end{figure*}

\section{Planck legacy}\label{sec:Planck}

\begin{figure*}
  \centering
  \includegraphics[width=1\textwidth]{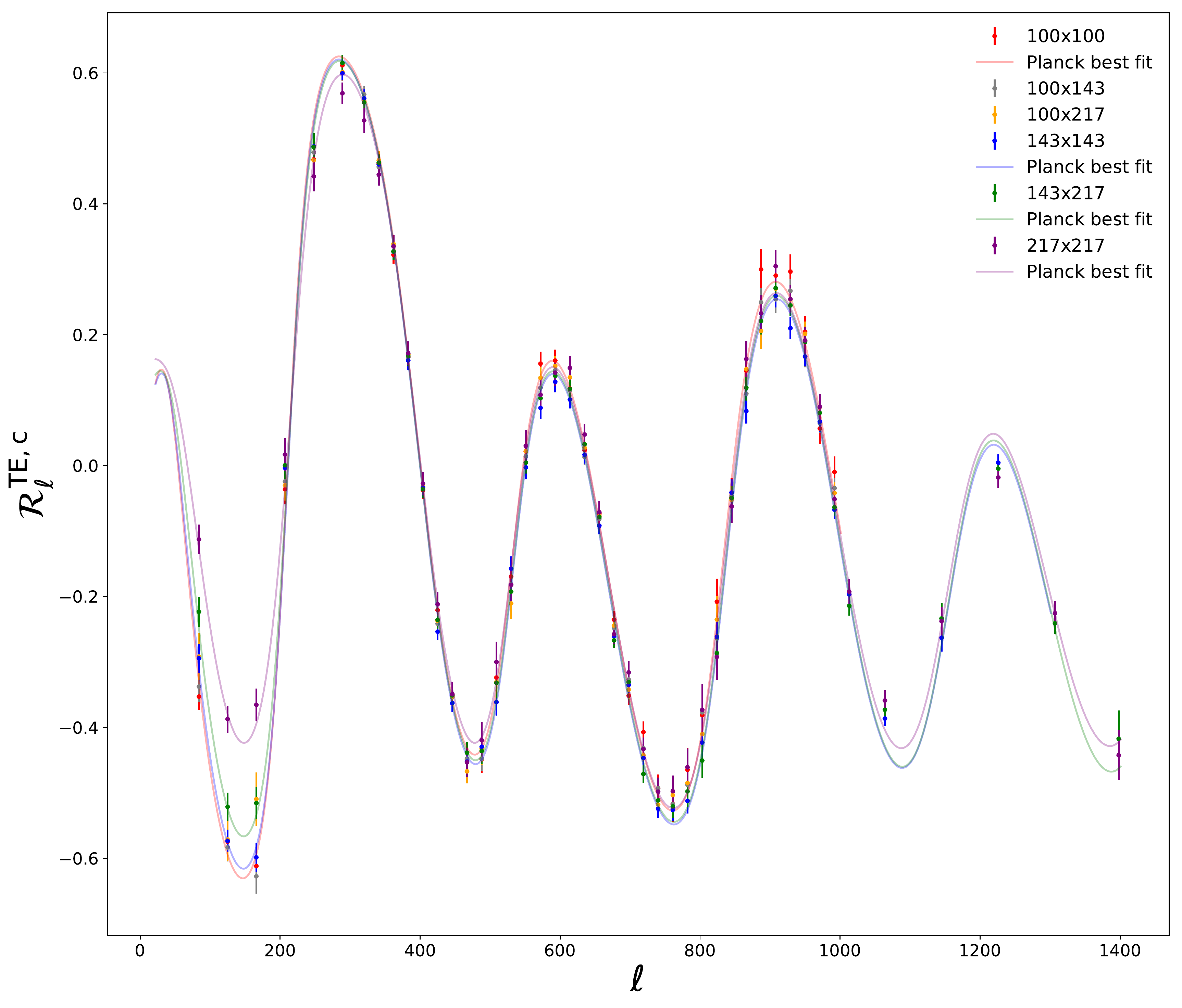}
  \caption{ All Planck legacy data correlation coefficients for  cross-frequency cross half-mission spectra, including the  100 GHz $\times$ 143 GHz and 100 GHz $\times$ 217 GHz spectra. The plik likelihood only models  the 100 GHz $\times$ 100 GHz, 143 GHz $\times$ 143 GHz, 143 GHz $\times$  217 GHz and 217 GHz $\times$ 217 GHz so we display the best-fit only for these spectra. The difference in the best-fits for different frequency pairs comes from foregrounds and temperature-to-polarisation leakage model. }
  \label{fig:all_cross}
\end{figure*}

In this section, we validate our calculation of the statistical properties of the correlation coefficient using simplified simulations of the Planck legacy survey. We then apply our results to the Planck legacy data and compare the resulting correlation coefficients of T and E modes with expectations from the $\Lambda$CDM model.

\subsection{Simulations}

In order to validate the analytic expression presented in Section \ref{sec:stat_properties}, we start by generating simplified Planck simulations. We generate gaussian realisations from  a set of lensed CMB power spectra also including systematics and foregrounds residuals\footnote{For each frequency pairs $\alpha, \beta$, we use the best-fit systematic and foregrounds curve publicly released by the Planck collaboration: {\scriptsize base\_plikHM\_TTTEEE\_lowl\_lowE\_lensing.minimum.plik\_foregrounds}, the power spectra used for the simulations are therefore given by $C^{X_{\alpha}  Y_{\beta}}_{\ell}= C^{\rm X Y,  CMB}_{\ell}+C^{ X_{\alpha} Y_{\beta}, \rm syst+foregrounds}_{\ell}$ with $X, Y = \{T, E\}$.}  and convolve them with the Planck beams computed over the fraction of the sky defined by the plik likelihood mask.
We note that plik do not provide best-fit residuals foregrounds for the $100 \ {\rm GHz} \times 143 \ {\rm GHz}$ and $100 \ {\rm GHz} \times 217 \ {\rm GHz}$ temperature power spectra. Obtaining them will require modifying the plik likelihood, or alternatively the use of a likelihood modelling all cross frequency spectra such as Hillipop \cite{2017A&A...606A.104C}. This is beyond the scope of the paper and we set the residual foreground model of these spectra to zero.  

We then add homogeneous gaussian noise based on the measured noise power spectra of the Planck data $ \hat{N}_{\ell}= \frac{1}{2}\left( \hat{C}^{\rm hm_{1} \times hm_{1}}_{\ell}+\hat{C}^{\rm hm_{2} \times hm_{2}}_{\ell}- 2 \hat{C}^{\rm hm_{1} \times hm_{2}}_{\ell} \right)$, and convolve the signal + noise simulation with the HEALPIX $N_{\rm side}=2048$  pixel window function. We do it for the two splits of data (half mission) of the three main cosmology channels: 100 GHz, 143 GHz, 217 GHz. We mask the maps using the plik likelihood masks and estimate the power spectra of the simulations $\hat{C}^{TT}_{\ell}, \hat{C}^{EE}_{\ell},\hat{C}^{TE}_{\ell}$ using the master algorithm implemented in \href{https://github.com/simonsobs/PSpipe}{PSPy}. We then bin the power spectra with a binning size ensuring that the estimated $\hat{C}^{EE}_{b}$ stay positive and further apply a multipole cut for each cross frequency spectra as indicated in Table \ref{tab:multipole_cut}. 

We  form estimates of the correlation coefficient $\hat{{\cal R}}^{\rm TE}_{b}= \hat{C}^{\rm TE}_{b}/\sqrt{  \hat{C}^{\rm TT}_{b} \hat{C}^{\rm EE}_{b}}$ and compare the bias and dispersion of the correlation coefficient of the simulations with our analytic expressions relating them to the $\hat{C}^{\rm TT}_{b},\hat{C}^{\rm EE}_{b},\hat{C}^{\rm TE}_{b}$ covariance matrix elements. The results are displayed on Figure \ref{fig:cov_and_bias}.
The bias is extremely small, or order  half a percent of ${\cal R}^{\rm TE}_{b}$ and measuring it precisely has required generating 1500 simulations of the Planck survey.  We find a satisfying agreement between the expectation for the statistical properties of $\hat{{\cal R}}^{TE}_{b}$ and the simulations.

Our noise simulations are simplistic and do not capture effects from the scanning strategy of Planck. This could have an impact on the computation of the covariance and the bias of the Planck legacy correlation coefficients.  In order to produce more realistic estimates of the statistical properties of the data, we replace our noise simulations by the 300 full focal plane noise simulations (FFP10) provided by the Planck collaboration \cite{2018arXiv180706207P}. These simulations include instrumental systematic effects and capture noise inhomogeneities from the Planck scanning strategy. In the following, we will use the errors derived from  the FFP10 simulations as a baseline for the errors on the data correlation coefficients.

\begin{table}
\caption{\small Multipoles cut used for Planck simulations and data}
\centering
\begin{tabular}{lcc}
\hline
\hline
     &  $\ell_{\rm min}$  &  $\ell_{\rm max}$  \\
\hline
100 GHz $\times$ 100 GHz & 50 & 1000 \\
100 GHz $\times$ 143 GHz & 50 & 1000  \\
100 GHz $\times$ 217 GHz  & 50 & 1000  \\
143 GHz $\times$ 143 GHz  & 50 & 1300  \\
143 GHz $\times$  217 GHz  & 50 & 1400  \\
217 GHz $\times$ 217 GHz & 50 & 1400  \\
\hline
\end{tabular}
\label{tab:multipole_cut}
\end{table}

\subsection{Data}

We then estimate $\hat{{\cal R}}^{\rm TE, c}_{b}$ of the Planck 2018 legacy data, and display the measurement of the correlation coefficients on Figure \ref{fig:results} and Figure \ref{fig:all_cross}. When available, we compare $\hat{{\cal R}}^{\rm TE, c}_{b}$ with the best-fit plik cosmology and foregrounds and compute the $\chi^{2}$ of the residuals
\ba
\chi^{2}= ( \hat{{\cal R}}^{\rm TE, c}_{b}- {\cal R}^{\rm TE, th}_{b})^{T} {\rm Cov}^{-1} ( \hat{{\cal R}}^{\rm TE, c}_{b}- {\cal R}^{\rm TE, th}_{b}).
\ea
We find that the $\chi^{2}$ of the measured correlation coefficients are consistent with expectations. 
The best-fit $\Lambda$CDM model obtained by fitting the three cross spectra $\hat{C}^{\rm TT}_{b},\hat{C}^{\rm EE}_{b},\hat{C}^{\rm TE}_{b}$ is also a good fit to $ \hat{{\cal R}}^{\rm TE, c}_{b}$, and no clear feature is visible in the residuals. While this result is expected as the same data are used in both cases, it is an interesting consistency test given the robustness of the correlation coefficient against multiplicative biases.  We make the resulting correlation coefficients public, they can be accessed at 
\href{https://github.com/simonsobs/PSpipe/tree/master/project/correlation_coeff/results}{\faGithub  }. An interesting extension of this work, left for a future study, would be to write a full likelihood for the correlation coefficients and estimate the cosmological parameters directly from them, this would allow to combine the different measurement of $ \hat{{\cal R}}^{\rm TE, c}_{b}$ and to check the consistency of the best-fit values of the cosmological parameters. 

\subsection{Gaussianity}
\begin{figure}
  \centering
  \includegraphics[width=1\columnwidth]{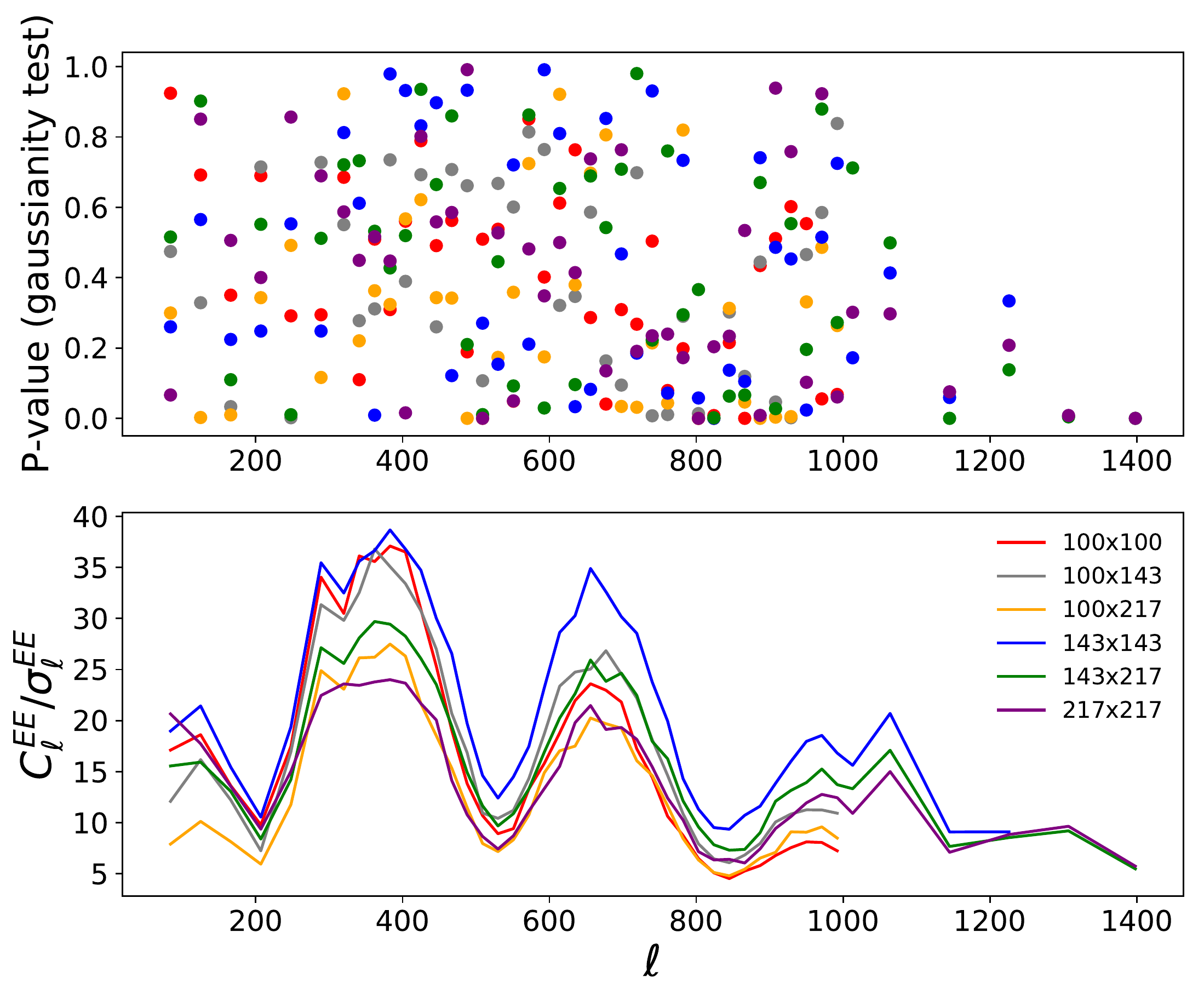}
  \caption{ Top: p-values of the null hypothesis that the binned correlation coefficients  are gaussian. Bottom: signal-to-noise of the binned EE power spectrum. We find that overall the distribution of the p-values is consistent with the null hypothesis that the distributions are gaussian. However, there are outliers with low p-values ($p<10^{-4}$) for the 250 tests that we have performed. They appear to correlate with the minima in signal-to-noise of the EE power spectra. }
  \label{fig:gaussianity}
\end{figure}
The correlation coefficients are formed as a ratio of power spectra, that are themselves computed as product of gaussian numbers. They therefore do not follow a gaussian distribution. However, from the central limit theorem, we expect that they converge to one in the limit of high multipoles and large enough bins. In order to test this hypothesis we used the Planck simulations to estimate the distribution of each bin values of the correlation coefficients. We then perform the D'Agostino and Pearson's test \cite{10.1093/biomet/60.3.613} that combines the skew and kurtosis of the distribution and results in a p-value for the null hypothesis that each bin distribution is gaussian. We report the results for all p-values in Figure \ref{fig:gaussianity}. We find that overall the distribution of the p-values is consistent with the null hypothesis that the distributions are gaussian. However, we find outliers with low p-values ($p<10^{-4}$) for the 250 tests that we have performed. They appear to correlate with the minima in signal to noise of the EE power spectra. To confirm it, we ran simulations with smaller noise level and found that all distributions become consistent gaussian. This reinforces the prescription that the correlation coefficients are easily defined only in the high signal to noise regime, and that a suitable bin size and multipole cut should be chosen for ensuring this requirement. We note that the skewness and kurtosis of the distributions could be computed analytically from  Eq \ref{eq:development}, we let this computation to future work.

\section{Discussion}\label{sec:discussion}

In this work, we have studied the TE correlation coefficient, a new observable of CMB physics insensitive to multiplicative bias arising from instrumental systematics. We have derived its variance and bias and shown that they can be computed from the  expression of the $C^{\rm TT}_{\ell},C^{\rm EE}_{\ell},C^{\rm TE}_{\ell}$ covariance matrix elements. We have then 
computed the correlation coefficients of Planck legacy data and we have found that they are consistent with expectations from the Planck best-fit LCDM and foregrounds model. 

Beyond its intrinsic physical meaning, the TE correlation coefficient represents a simple and powerful projection of the data in a space orthogonal to most systematics affecting CMB experiments. It allows for simple visualisation of residuals with respect to the best-fit cosmological model.

While fitting simultaneously for the $C^{\rm TT}_{\ell},C^{\rm EE}_{\ell},C^{\rm TE}_{\ell}$ power spectra, an implicit fit of the correlation coefficient is made, we believe that the discussion in this paper is reinforcing the importance of precise measurements of the polarisation E mode anisotropies, since they allow to break the degeneracy between parameters affecting the overall amplitude and shape of the spectra and parameters affecting the position and relative amplitude of the peaks.

The next generation of CMB experiments such as Simons Observatory \cite{2019JCAP...02..056A}  and CMB S4 \cite{2016arXiv161002743A}  will produce cosmic variance limited measurements of E modes for a wide range of angular scales. This will result in very precise measurements of the TE correlation coefficient. An interesting follow up study will be to precisely quantify how constraining the correlation coefficient is on the extensions to the $\Lambda {\rm CDM}$ model, in particular on extensions changing the acoustic scale by changing physics near the time of recombination.

Finally, while we have focused in this paper on the correlation coefficient that can be formed using temperature and E modes measurement, the results on the statistical properties of ${\cal R}^{\rm TE}_{b}$ could be generalised and applied on large scale structure data, in particular for the correlation coefficient between density and lensing convergence fluctuations that are strongly affected by multiplicative biases.

\section*{Acknowledgments}
TL and ZL thanks Jo Dunkley, Erminia Calabrese, Silvia Galli, Karim Benabed and Antony Lewis for interesting discussions on the Planck legacy data. TL thanks David Spergel for discussion about the gaussianity of the estimator and Sigurd Naess for interesting discussion in general.
Some of the results in this paper have been derived using the  \href{https://healpix.sourceforge.io/downloads.php}{HEALPix} \cite{2005ApJ...622..759G} package. 
This research used resources of the National Energy Research Scientific Computing Center (NERSC), a U.S. Department of Energy Office of Science User Facility operated under Contract No. DE-AC02-05CH11231.

 \bibliography{draft}

\end{document}